\documentclass[conference]{IEEEtran}
\IEEEoverridecommandlockouts
\usepackage{cite}
\usepackage{amsmath,amssymb,amsfonts}
\usepackage{algorithmic}
\usepackage{graphicx}
\usepackage{textcomp}
\usepackage{xcolor}
\usepackage{array}

\def\BibTeX{{\rm B\kern-.05em{\sc i\kern-.025em b}\kern-.08em
    T\kern-.1667em\lower.7ex\hbox{E}\kern-.125emX}}
\begin{document}

\title{Bimorph Lithium Niobate Piezoelectric Micromachined Ultrasonic Transducer}

\author{
\IEEEauthorblockN{Ziqian Yao, Vakhtang Chuluhadze, Zihuan Liu, Xiaoyu Niu, Tzu-Husan Hsu, Byeongjin Kim, Neal Hall, \\and Ruochen Lu}

\IEEEauthorblockA{Electrical and Computer Engineering,
University of Texas at Austin, Austin, TX}
}

\maketitle

\begin{abstract}
This work demonstrates a prototype bi-layer piezoelectric micromachined ultrasonic transducer (PMUT) based on transferred periodically poled piezoelectric film (P3F) X-cut lithium niobate (LN). Opposite in-plane polarizations in the piezoelectric film stack are employed to enable efficient lateral field excitation of the flexural mode. Thanks to its high piezoelectric coefficient and low dielectric loss, the X-cut LN exhibits high figure of merits (FoMs) as both sensors and transducers. The fabricated PMUT demonstrates an out-of-plane mode near 1 MHz with an electromechanical coupling of 3.6\%. Laser Doppler vibrometry further validates the finite element analysis, showing a peak center displacement of 8nm /V. These results establish bi-layer P3F LN PMUTs as a promising platform for compact and high-performance ultrasonic transducers. Future work will focus on theoretical analysis, modeling of the measured data, improving the design of the transducer topology, and mitigating feedthrough effects. 

\end{abstract}


\section{Introduction}

Piezoelectric micromachined ultrasonic transducers (PMUTs) enable compact ultrasonic sensing and actuation using flexural vibration of thin piezoelectric films. Unlike capacitive micromachined ultrasonic transducers (CMUTs), PMUTs eliminate DC bias requirements and offer linear electromechanical conversion \cite{helmerich2025multiple, moisello2024pmut}, allowing simpler drive electronics and higher output at low voltages. Advancements in thin-film platforms such as sputtered AlN/ScAlN \cite{akiyama2009influence,wingqvist2010increased,lu2018surface,wall2022sputtering}, and PZT \cite{liu2019characterization,lin2009pzt,thongrit2023improving} have enabled PMUT applications in medical imaging \cite{liu2022fabrication,atheeth2023review}, wireless communications \cite{pop2018novel,wang2018wireless}, fingerprint recognition \cite{jiang2017monolithic,jiang2017ultrasonic}, and flow sensing \cite{seo2019piezoelectric,xiu2022development}. However, achieving strong performance in both transmission and reception remains a key challenge for PMUTs, largely limited by the electromechanical properties of the piezoelectric material. Recent efforts have explored piezoelectric materials with improved coupling and lower acoustic loss \cite{muralt2017best, xia2024high,zhao2025piezoelectric}.

To enable fair comparison between piezoelectric materials regardless of transducer geometry, transmitter and receiver figure-of-merits (FoMs) have been developed~\cite{muralt2017best}. Table~\ref{tab:SensorTransducerFoM} summarizes these FoMs for widely used PMUT materials, including AlN, ScAlN, and PZT. To characterize the electrical output generated by mechanical deformation in flexural transducers, several figure-of-merits (FoMs) are defined.

The piezoelectric coefficient \( e \) (C/m\(^2\)) quantifies the induced surface charge per unit stress, representing current and charge sensitivity. In contrast, the ratio \( e/\varepsilon \) (GV/m) reflects voltage sensitivity by accounting for dielectric constants from the material’s permittivity \( \varepsilon \). To evaluate detection performance under low-signal conditions, the FoM \( e/\sqrt{\varepsilon \tan\delta} \), where \( \tan\delta \) is the dielectric loss tangent. A higher value indicates improved signal-to-noise ratio (SNR) in receiver-limited systems. For transducer applications, a large electromechanical coupling coefficient (\(k^2\)) is essential to maximize energy conversion between electrical and mechanical domains. In flexural-mode devices with a bending layer, this efficiency is quantified by the transducer FoM \(e^2/(\varepsilon Y)\), which captures bidirectional energy transfer. Here, \(Y\) represents the stiffness of the passive layer, normalized to the Young’s modulus of silicon \cite{muralt2017best}. Among the materials listed in Table~\ref{tab:SensorTransducerFoM}, PZT and PMN-PT exhibit high \(k^2\), making them favorable for transducer applications. However, their large dielectric constant and high dielectric loss significantly degrade their performance as sensors, leading to poor voltage sensitivity and high noise levels. This trade-off highlights the need to explore alternative piezoelectric materials with balanced properties that support strong electromechanical coupling while maintaining low dielectric losses, which enables improved performance in both sensing and transmitting modes.

Lithium niobate (LN), with its high piezoelectric coefficients and low dielectric loss \cite{gong2013figure, bhugra2017piezoelectric}, has emerged as a strong candidate for next-generation PMUTs. Previous work \cite{lu2020piezoelectric} has demonstrated promising PMUT performance using single-layer 36°Y-cut LN under lateral-field-excitation (LFE). Building on this, this work introduces a multilayer periodically poled piezoelectric film (P3F) structure, fabricated through the high-quality transfer of single-crystal LN. By utilizing a bimorph X-cut LN structure with alternating crystal orientations, this approach fully harnesses LN’s advantageous FoMs for both sensing and actuation.

\begin{table}[!t]
\caption{Sensor and Transducer FoMs of Piezoelectric Materials}
\label{tab:SensorTransducerFoM}
\centering
\scriptsize
\setlength{\tabcolsep}{3.5pt}  
\renewcommand{\arraystretch}{1}  
\begin{tabular}{|c|c|c|c|c|c|}
\hline
\textbf{Material} & \textbf{Ref.} & 
\multicolumn{3}{c|}{\textbf{Sensor FoMs}} &
\textbf{Transducer FoM} \\
\cline{3-5}
& & 
$e$ (C/m$^2$) & 
$e/\varepsilon$ (GV/m) & 
\shortstack[l]{$e/\sqrt{\varepsilon \tan \delta}$\\(10$^6$ (J/m$^3$)$^{1/2}$)} &
$e^2/(\varepsilon Y)$ \\
\hline
PZT         & \cite{baek2011giant} & 18.7  & 1.30  & 1.10 & 0.19 \\
PMN-PT      & \cite{baek2011giant} & 26.0  & 1.90  & 0.84 & 0.38 \\
AlN         & \cite{bhugra2017piezoelectric} & 0.68  & 6.99  & 2.20 & 0.04 \\
Sc$_{0.3}$AlN & \cite{mertin2017enhanced} & 2.25  & 12.7  & 3.10 & 0.22 \\
Y-LN        & \cite{wong2002properties} & 2.43  & 6.24  & 3.43 & 0.04 \\
X-LN        & \cite{wong2002properties} & 4.65  & 13.13 & 6.85 & 0.47 \\
\hline
\end{tabular}
\end{table}

\section{Design and Simulation}

\begin{figure}[!t]
  \centering
  \includegraphics[width=\linewidth]{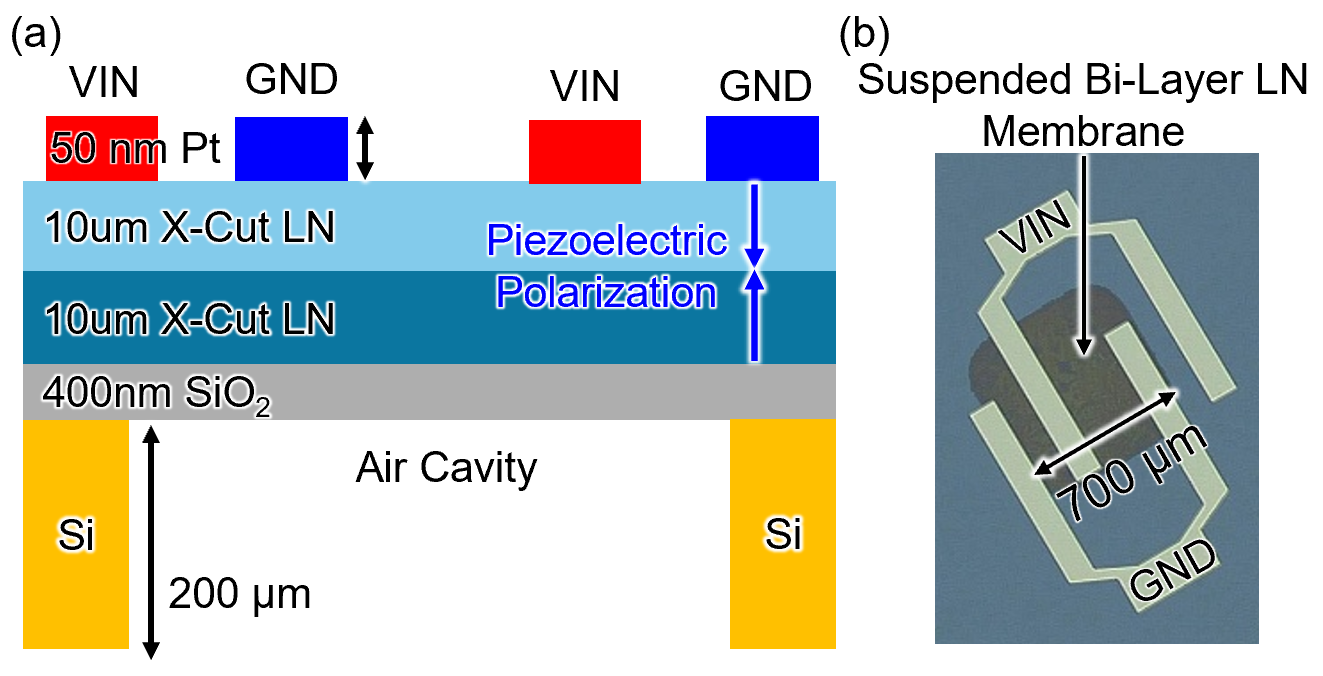}
  \caption{(a) Top view cross-section of the lateral field excited bimorph X-cut LN PMUT. (b) Optical image of the fabricated suspended LN PMUT with top electrodes labeled as VIN and GND.}
  \label{fig1}
\end{figure}

The proposed bimorph X-cut PMUT top view is shown in Fig. \ref{fig1} with key parameters labeled. As shown in Fig.~\ref{fig1}(a), the device is composed of a suspended trilayer stack consisting of two 10~$\mu$m thick X-cut LN layers and a 400~nm SiO$_2$ interlayer on top of a Si carrier wafer. The bottom LN layer is rotated in-plane by 180$^\circ$, forming a P3F structure that reverses the polarization axis (Z) orientations. This configuration promotes constructive charge buildup across the bimorph stack while suppressing undesired overtones through charge cancellation from opposing stress phases, thereby improving out-of-plane actuation \cite{yao2025periodically,naumenko2024enhancement}. Two pairs of 100~nm platinum (Pt) electrodes are patterned on the top surface for flexural mode excitation. The underlying Si is removed beneath the active region to create a suspended cavity, providing mechanically free boundaries for acoustic displacement. Fig.~\ref{fig1}(b) shows an optical image of the fabricated device, where a $700~\mu\text{m} \,\times\, 700~\mu\text{m}$ cavity region is clearly visible with a suspended membrane on top, along with the patterned metal electrodes labeled as VIN and GND.

\begin{figure}[!t]
  \centering
  \includegraphics[width=\linewidth]{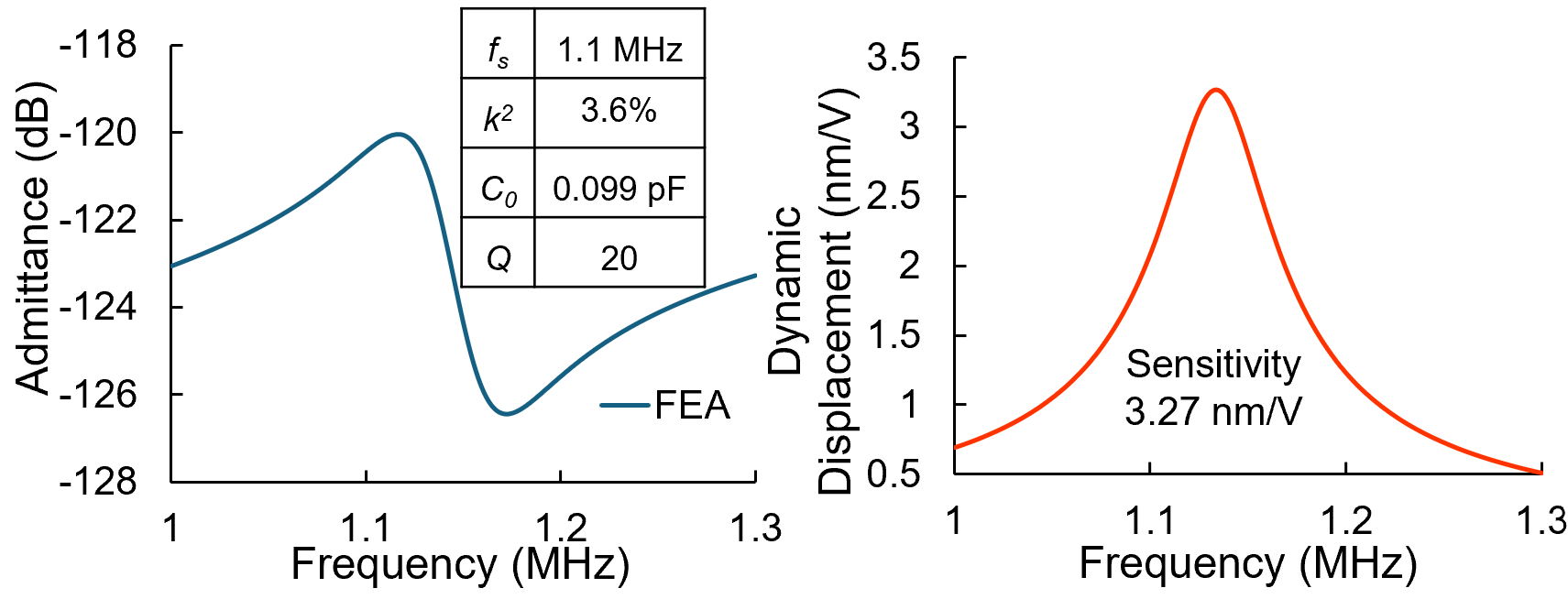}
  \caption{Finite-element simulation of the bimorph X-cut LN PMUT: (a) 3D displacement mode shape under lateral-field excitation, (b) cross-sectional stress distribution showing constructive polarization, (c) simulated electrical admittance with extracted resonance parameters, and (d) dynamic out-of-plane displacement near resonance, demonstrating a sensitivity of 3.27 nm/V}
  \label{fig2}
\end{figure}

The device structure is first analyzed using COMSOL finite element analysis (FEA). As shown in Fig. \ref{fig2}(a–b), a lateral electric field applied across the top electrodes induces in-plane stress ($T_x$) in the LN layers through the strong $e_{11}$ coefficient (4.65 C/m$^2$). Because the bimorph LN stack has reversed Z-axis orientations, the two layers generate stresses of opposite sign under the same field, which add constructively across the thickness to enhance out-of-plane displacement. The film stack, consisting of 10 µm LN and 2~$\mu$m SiO$_2$, is designed to prevent charge cancellation and maximize electromechanical coupling. Electrodes are alternately biased and placed near stress antinodes, while the outer electrodes extend beyond the suspended region to improve energy transfer. The simulated admittance spectrum in Fig. \ref{fig2}(c) shows a clear resonance around 1.1 MHz, corresponding to an effective coupling of 3.6 \%, which is higher than reported ScAlN counterparts \cite{wang2017design, wang2016scandium}. The static capacitance ($C_0$) is extracted as 0.099 pF and the quality factor ($Q$) is set to be 20. Meanwhile, Fig. \ref{fig2}(d) plots the dynamic displacement, peaking at 3.27 nm/V near resonance, highlighting the strong out-of-plane vibration.

\begin{figure}[!t]
  \centering
  \includegraphics[width=\linewidth]{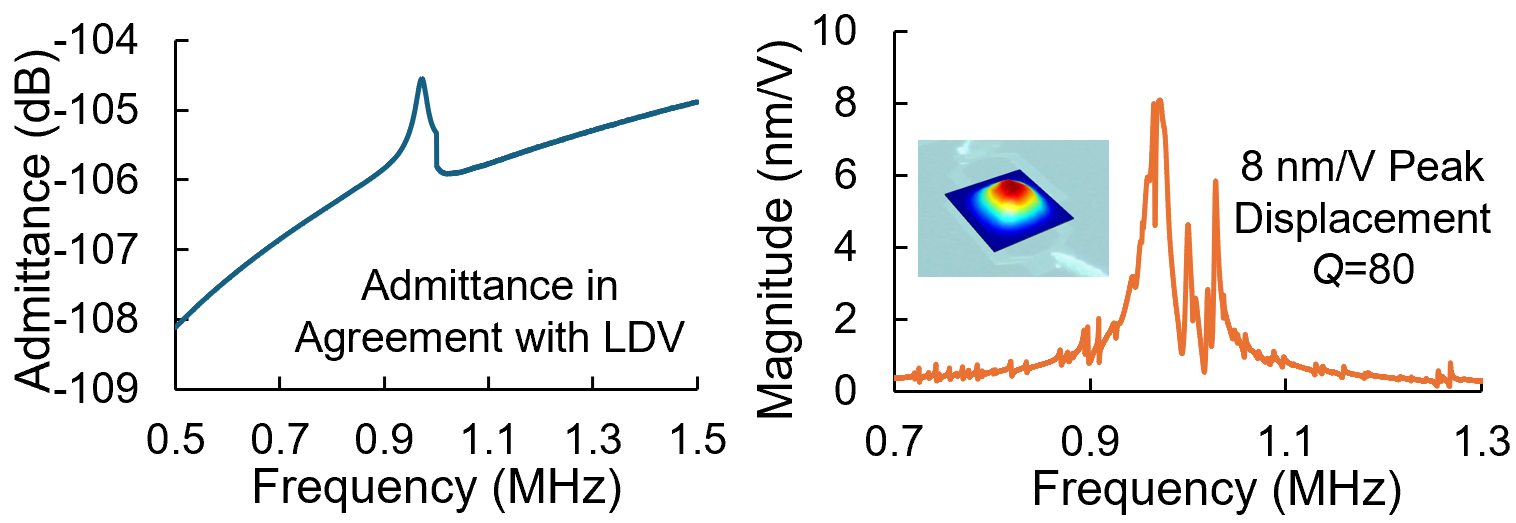}
  \caption {(a) Measured electrical admittance of the bimorph X-cut LN PMUT. (b) Out-of-plane diaphragm displacement measured by LDV, showing a resonance near 0.97 MHz with a peak magnitude of 8 nm/V.}
  \label{fig3}
\end{figure}

\section{Fabrication and Measurement}
The fabricated bimorph LN PMUT is shown in Fig.~\ref{fig1}(b). The LN–SiO$_2$–Si film stack is provided by NGK Insulators. Fabrication begins with patterning 100-nm Pt top electrodes, followed by structural release via backside deep reactive ion etching (DRIE) of the silicon substrate. The etching was terminated at the buried oxide layer to avoid over-etching into the LN active layer. During processing, a slight backside-alignment offset was observed between the intended diaphragm window and the etched cavity. This offset resulted in a minor shift in the effective diaphragm dimensions, which subsequently caused a small change in operating frequency compared to FEA-simulated results. Future optimization will focus on optimizing the etch window and electrode tolerances. 

The fabricated devices were characterized in air using a laser Doppler vibrometer (LDV) to measure diaphragm displacement directly. Electrical admittance measurements, performed using the Zurich impedance analyzer and presented in Fig. 3(a), confirm the presence of a resonance at approximately 0.97 MHz. The minor discontinuity is caused by a switch in measurement range using the impedance analyzer. The higher admittance than FEA indicates feedthrough which will be studied in future works. The corresponding out-of-plane diaphragm displacement spectrum, shown in Fig. \ref{fig3}(b), exhibits a sharp resonance peak with a measured displacement of approximately 8 nm/V. The reduced displacement is potentially caused by the substrate feedthrough parasitics. The measured resonance is slightly down-shifted from FEA predictions of 1.1 MHz. This downshift is expected due to electrode mass loading and parasitics. Nevertheless, the measured maximum diaphragm displacement of 340 pm/V agrees with the expected mode shape and validates the bimorph actuation mechanism in Fig. \ref{fig2}.

\section{Conclusion}
In this work, we have demonstrated a prototype bimorph PMUT based on periodically poled piezoelectric film (P3F) LN. By leveraging opposite in-plane polarizations in an X-cut LN bimorph stack, we enabled efficient lateral-field excitation of the flexural mode and achieved strong electromechanical coupling. These results establish bi-layer P3F LN PMUTs as a promising platform for compact and high-performance ultrasonic transducers, offering a pathway toward next-generation sensing and actuation applications. Future work will focus on modeling of the measured data, improving the design of the transducer topology, and mitigating feedthrough effects.

\section{Ackowledgement}
The authors would like to thank DARPA HOTS for funding support and Dr. Todd Bauer for helpful discussions.

\renewcommand{\baselinestretch}{0.95}\renewcommand{\baselinestretch}{1.0}\normalsize

\end{document}